# Owi: Performant Parallel Symbolic Execution Made Easy, an Application to WebAssembly

Léo Andrès[a,c] , Filipe Marques[b] , Arthur Carcano[a] , Pierre Chambart[a] , José Fragoso Santos[b] , and Jean-Christophe Filliâtre[c]

a  OCamlPro SAS, 21 rue de Châtillon, 75014, Paris, France
b  INESC-ID / Instituto Superior Tecnico, University of Lisbon, Rua Alves Redol 9, 1000−029, Lisbon, Portugal
c  Université Paris-Saclay, CNRS, ENS Paris-Saclay, Inria, Laboratoire Méthodes Formelles, 91190, Gif-sur-Yvette, France

**Abstract**    In this paper, we present the design of Owi, a symbolic interpreter for WebAssembly written in OCaml, and how we used it to create a state-of-the-art tool to find bugs in programs combining C and Rust code. WebAssembly (Wasm) is a binary format for executable programs. Originally intended for web applications, Wasm is also considered a serious alternative for server-side runtimes and embedded systems due to its performance and security benefits. Despite its security guarantees and sandboxing capabilities, Wasm code is still vulnerable to buffer overflows and memory leaks, which can lead to exploits on production software. To help prevent those, different techniques can be used, including symbolic execution.

Owi is built around a modular, monadic interpreter capable of both normal and symbolic execution of Wasm programs. Monads have been identified as a way to write modular interpreters since 1995 and this strategy has allowed us to build a robust and performant symbolic execution tool which our evaluation shows to be the best currently available for Wasm. Moreover, because WebAssembly is a compilation target for multiple languages (such as Rust and C), Owi can be used to find bugs in C and Rust code, as well as in codebases mixing the two. We demonstrate this flexibility through illustrative examples and evaluate its scalability via comprehensive experiments using the 2024 Test-Comp benchmarks. Results show that Owi achieves comparable performance to state-of-the-art tools like KLEE and Symbiotic, and exhibits advantages in specific scenarios where KLEE's approximations could lead to false negatives.

# The Art, Science, and Engineering of Programming

**Owi: Performant Parallel Symbolic Execution Made Easy, an Application to WebAssembly**

# 1 Introduction

WebAssembly (Wasm) [29] is a binary compilation target designed to become the new standard for the web. Supported by almost all web browsers [12], Wasm enables web applications to benefit from improved performance and security. Beyond the web, Wasm powers server-side runtimes [1, 8, 16], IoT platforms [30], and blockchain smart contracts [33], extending its impact far beyond traditional web use cases.

By allowing high-level programming languages such as C, C++, Rust, Java, Go, Haskell, and OCaml to compile to Wasm, new classes of bugs and vulnerabilities are introduced into web and non-web applications alike. Issues in source code — such as buffer overflows [41], format string vulnerabilities [4], and memory leaks [27] — are propagated into Wasm binaries via compilation [37]. These vulnerabilities can be exploited to launch attacks such as cross-site scripting (XSS), denial of service (DoS), and code injection. Hence, it is crucial to provide developers and security analysts with robust tools to analyze and verify the security of Wasm binaries.

Symbolic execution [36], a powerful technique that explores all feasible execution paths of a program by using symbolic inputs instead of concrete ones, has been successfully applied to various programming languages, including C [9, 26], Java [46], and JavaScript [24, 25, 48]. Recently, symbolic execution tools for Wasm, such as WANA [33], Manticore [43], SeeWasm [31], and WASP [40], have emerged and represent the current state-of-the-art. However, these tools are either not fully-automatic, need deep knowledge of Wasm binaries, or tend to scale poorly on larger codebases [40].

To address these challenges, we present Owi, a toolkit for working with Wasm in the OCaml ecosystem. Owi includes a reference interpreter for Wasm capable of both concrete and symbolic execution. In this paper, we first describe how we developed reusable components and a modular interpreter from a concrete one, enabling shared code between the concrete and symbolic interpreters (Section 3). We then detail our implementation of a multi-thread choice monad (Section 4) based on a cooperative coroutine scheduler. Next, we show how **Owi can be used to perform symbolic execution, of Wasm, C, Rust and cross-languages programs** (Section 5). Finally, we present in (Section 6) the results of our experimental evaluation showing that **for Wasm, Owi outperforms all prior work tools** with an average speedup of 313× over Manticore [43], 57× over SeeWasm [31], and 4× over WASP [40], and **for C, Owi has results comparable to the state of the art software testing tools**, such as KLEE [9] and Symbiotic [14].

**Contributions** In summary, our contributions are as follows: **(1) Owi**: a toolkit to work with Wasm within the OCaml ecosystem, featuring a monadic reference interpreter for Wasm (Section 3); **(2) Scalable symbolic execution**: A robust symbolic execution engine for Wasm using a multi-core choice monad (Section 4); **(3) Cross-language bug-finding**: A front-end for symbolic execution of C and Rust programs via Wasm compilation (Section 5.2).

```
1  (module
2    ;; This function does nothing from the point of view
3    ;; of its caller but internally swap x and y if needed
4    ;; so that x < y and assert that the swap happened correctly
5    (func $test_swap (param $x i32) (param $y i32)
6      ;; if x > y
7      (if (i32.gt_s (local.get $x) (local.get $y))
8      (then
9        ;; Swap x and y using integer arithmetic
10       ;; and not a temporary variable
11       ;; x <- x + y
12       (local.set $x
13         (i32.add (local.get $x) (local.get $y)))
14       ;; y <- x - y
15       (local.set $y
16         (i32.sub (local.get $x) (local.get $y)))
17       ;; x <- x - y
18       (local.set $x
19         (i32.sub (local.get $x) (local.get $y)))
20       ;; We entered this branch with x > y and should
21       ;; have swapped the values, so check that now
22       ;; x < y by raising an exception if x - y > 0
23       (if (i32.gt_s
24         (i32.sub (local.get $x) (local.get $y))
25         (i32.const 0))
26       (then
27           ;; raise en exception (ie. "trap")
28           unreachable
29       ))
30     ))
31   )
32 )
```

■ **Listing 1** Example Wasm program written in Wasm Textual Format (Wat). This is a Wasm translation of a code example from [35] originally in C.

## 2 Background

This section provides a brief overview of Wasm, focusing on its syntax and semantics (Section 2.1), followed by an introduction to symbolic execution, with particular emphasis on the method applied in this paper (Section 2.2).

### 2.1 Wasm

Wasm [29, 47] is a low-level binary instruction format designed for portability, performance, and security. It provides a compact representation, efficient validation, and fast compilation while ensuring safe execution with minimal overhead. Wasm is platform-agnostic, being independent of specific hardware architectures, programming languages, or runtime environments. It is predominantly used as a compilation target for high-level languages and is deployed in web browsers, cloud platforms, and standalone runtimes.

A Wasm program is encapsulated in a module, which includes a collection of Wasm functions, shared global variables, and a specification for a linear memory (a global byte array acting as the program's heap). Wasm employs a stack machine model of computation, where values are pushed to and popped from the stack during execution. A Wasm module runs within an embedder, a host engine responsible for loading modules, resolving imports and exports between them, managing I/O, and handling exceptions such as traps.

The syntax of Wasm programs is illustrated in Listing 1, showcasing key elements such as functions (e.g., `$test_swap`), local variables (*e.g.*, `$x` and `$y`), constants (e.g., `i32.const 0`), and instructions (e.g., `i32.add`, `local.get`, `if`, `unreachable`). Wasm supports four primitive types: 32-bit integers (`i32`), 64-bit integers (`i64`), single-precision floating-point numbers (`f32`), and double-precision floating-point numbers (`f64`). Wasm instructions determine whether integer operations are signed or unsigned through a sign extension. Wasm variables are either local or global, with local variables existing within a function's scope and global variables shared across the module.

The primary memory is structured as a large, linear array of bytes known as linear memory. This memory can be shared between modules through imports and exports, and although its initial size is fixed, it can be programmatically grown when needed. In contrast to typical stack machines, Wasm supports structured control flow with constructs such as `if`, `else`, and `loop`, making Wasm code more readable and easier to reason about for developers.

For a detailed treatment of Wasm's syntax and semantics, readers are referred to [29].

### 2.2 Symbolic Execution

Symbolic execution is a program analysis technique used to explore all feasible execution paths of a program up to a given bound [36]. Instead of using concrete inputs, symbolic execution operates on symbolic inputs, which represent multiple possible values. Each time the symbolic execution engine encounters a conditional branch whose condition depends on the value of one or several symbols, it forks the execution, exploring both possible branches. For each path, the engine builds a logical formula called the path condition, representing the constraints on the inputs needed to reach that path.

When symbolic execution encounters a branch, the path condition is updated with the corresponding guard for the then branch or the negation of the guard for the else branch. To check the feasibility of these paths and validate assertions within the program, symbolic execution engines rely on an underlying SMT (Satisfiability Modulo Theories) solver like Z3 [20], Colibri2 [6], Bitwuzla [44], Alt-Ergo [17] or cvc5 [3]. An execution path is said to be feasible if the SMT solver can find at least one concrete set of inputs that satisfies the path condition. Additionally, an assertion

L. Andrès, F. Marques, A. Carcano, P. Chambart, J. Fragoso Santos, J.-C. Filliâtre

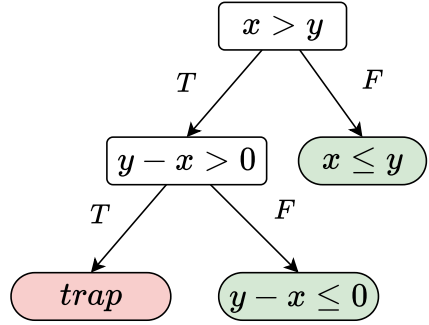

**Figure 1** Execution tree of the `$test_swap` function. Symbolic execution explores exhaustively all these execution paths. When reaching the trap node, the SMT provides the values leading to that node.

at a particular point in the program holds if it is implied by the path condition at that point.

**Symbolic Execution Example** To illustrate symbolic execution in practice, consider the `$test_swap` function in the Wasm program shown in Listing 1. In one path, the function ends with an `unreachable` instruction, which should never be executed under any input and raises an exception if it is. Symbolic execution is used to verify this by exhaustively exploring all feasible execution paths, as depicted in the execution tree in Figure 1. In the figure, green leaf nodes represent valid execution outcomes, while red nodes indicate paths that trigger the `unreachable` instruction.

Consider inputs $x = 8388481$ and $y = -2147483648$, which cause 32-bit integer overflow during the subtraction $x - y$ in the second branch (*i.e.* $-2147483648 - 8388481 \mod 2^{32} = 2139095167$), leading to the `unreachable` trap. Below, we explain how this input can be discovered.

Since there are three execution paths through the `$test_swap` function, each must be symbolically executed. For this example, we assume a breadth-first exploration strategy. Initially, the symbolic execution assigns symbolic values to the local variables $x$ and $y$. Upon encountering the first conditional branch, execution forks into two paths: one where $x > y$, and the other where $x \leq y$. The path conditions are updated accordingly:

$$\text{PC}_T \equiv x > y \quad \text{and} \quad \text{PC}_F \equiv x \leq y$$

Execution then proceeds along the path described by $\text{PC}_T$. In this path, the values of $x$ and $y$ are swapped, and the second conditional branch (`$x - $y > 0`) is encountered, generating two new path conditions:

$$x' \mapsto x + y \quad y \mapsto x' - y \quad x'' \mapsto x' - y$$

$$\text{PC}_{TT} \equiv x > y \land y'' - x'' > 0 \quad \text{and} \quad \text{PC}_{TF} \equiv x > y \land y'' - x'' \leq 0$$

Both path conditions are feasible, leading to another fork in the execution. Eventually, $\text{PC}_{TT}$ reaches the `unreachable` instruction. An SMT solver can be queried to find concrete inputs for $\text{PC}_{TT}$, yielding:

$$\$x \mapsto 8388481 \quad \$y \mapsto -2147483648$$

This demonstrates the power of symbolic execution in identifying edge cases, such as integer overflow, that can cause unexpected behaviors. Although further paths, such as $\text{PC}_{TF}$, remain to be explored, the discovery of an erroneous path already demonstrates the utility of the analysis.

## 3 Turning a Concrete Interpreter to a Monadic One

Owi is an open-source[1] Wasm toolkit developed in OCaml. Its interpreter fully complies with the Wasm standard [29] and also supports some of its recent standard and non-standard extensions.[2] In this section, we present how we have turned Owi, historically a robust reference interpreter for Wasm, into a modular and general interpreter able to do both concrete and symbolic execution.

This section is organized as follows: first, we provide an overview of the implementation of the concrete interpreter (Section 3.1). Next, we describe how to parametrize the interpreter over different implementations of its base values (Section 3.2). Finally, we show how to parametrize the interpreter over its evaluation strategy, implemented as a monad, which allows us to turn it into a full symbolic execution engine (Section 3.3).

### 3.1 The Concrete Interpreter

The code of the concrete interpreter is modeled on the semantics of Wasm. For the sake of simplicity, we focus in this article on a reduced Wasm syntax that only has integer types.

It is represented by the following OCaml algebraic data types:

```
type value = I32 of int32 | I64 of int64
```

In this representation, values of the `value` type represent a 32- or 64-bit Wasm integer. An OCaml value representing the Wasm 32-bit integer 42 is written as `I32 42l`, where the `l` indicates that this is a 32-bit integer and not OCaml's default integer with 63 or 31 bits.

---

[1] Owi is available under the AGPL-3.0 license at https://github.com/ocamlpro/owi (visited on 2024-10-22).

[2] Owi currently supports the following standard extensions: Import/Export of Mutable Globals, Non-trapping float-to-int conversions, Sign-extension operators, Multi-value, Reference Types, and Bulk memory operations; and not yet standard extensions: Tail call, Typed Function References, and Extended Constant Expressions.

```ocaml
 1 let select x =
 2   match x with I32 x when x <> 0l -> true | _ -> false
 3
 4 let eval_instr i stack =
 5   match (i, stack) with
 6   | Binop Add, I32 x :: I32 y :: stack ->
 7     I32 (Int32.add x y) :: stack
 8   | Binop Gt, I32 x :: I32 y :: stack ->
 9     I32 (Int32.gt x y) :: stack
10   | If (if_true, if_false), cond :: stack ->
11     let cond = select cond in
12     if cond then eval if_true stack
13             else eval if_false stack
```

■ **Listing 2** Evaluator for the simplified Wasm instructions. We omit the cases that would result in Wasm type errors (e.g. an empty stack).

The instructions are also defined as an OCaml datatype, illustrated by the following subset:

```ocaml
type binop = Add | Gt
type instr = Binop of binop | If of (instr list * instr list)
```

Instructions can either be a binary operation `Binop op`, where op can be `Add` or `Gt`, or a conditional `If (if_true, if_false)`, where if_true (resp. if_false) is the instruction list to be executed if the condition at the top of the stack is true (resp. false).

In OCaml, type signatures specify tuples using the `*` operator. When constructing tuples, pair notation is used, such as (1, 2), to represent a tuple of type `int * int`.

We implement an interpreter for this syntax by defining an evaluator for Wasm *blocks* (lists of instructions). The evaluator is a function with the following signature:

```ocaml
val eval : instr list -> value list -> value list
```

It takes an `instr list` (representing a block of instructions) and a `value list` (representing Wasm's value stack) as arguments and returns a `value list` (the modified stack after evaluation).

Importantly, note that the `->` symbol in OCaml represents a *function type arrow*, which specifies the types of a function's input and output. This is due to OCaml's use of *currying*. For more information on OCaml type signatures, readers can refer to [39].

This block evaluator is built on a single instruction evaluator, that also updates the stack accordingly. It has the following signature and its implementation is shown in Listing 2.

```ocaml
val eval_instr : instr -> value list -> value list
```

The `eval_instr` function pattern matches on `(i, stack)`, where i is the instruction to evaluate, and stack is a list of values. The key cases are as follows:

1. `Binop Add`: The OCaml operator `::` is the list consing operator, so `I32 x :: I32 y :: stack` represents a list whose first element is `I32 x`, second element is `I32 y` and the tail of the list is a list bound to the variable `stack`. This pattern matches a stack with two 32-bit integers at its top, `I32 x` and `I32 y`. The result of the addition is calculated using OCaml's `Int32` module, which provides operations on 32-bit integers, such as `Int32.add` for addition. A new stack is then returned with the addition result at its start followed by the tail.
2. `Binopt Gt`: Similarly to the previous case, two integers are "popped" from the stack, `I32 x` and `I32 y`. We perform a *greater than* comparison on them and the result is "pushed" as a truthy value (1 for true, 0 for false).
3. `If (if_true, if_false)` with a value `cond` at the top of the stack. If `cond` is truthy (*i.e.*, `cond` is `I32 x` where $x \neq 0$), the `if_true` block is executed; otherwise, the `if_false` block is executed.

### 3.2 The Parametric Interpreter

To generalize the concrete interpreter, we introduce a parametric version by abstracting over value types. This is achieved using an OCaml *functor*, which allows us to parameterize the interpreter over different modules representing the value types. In OCaml, a module is a collection of values or functions grouped together under a namespace. A functor is similar to a function but operates on *modules* and not values: it is a function at the module-level.

Listing 3 shows the `Interpreter` functor, it takes as parameter a module `Value` that defines operations for the types representing Wasm values. The interpreter then uses these operations, allowing us to replace the concrete `Int32` and `Int64` operations with abstract counterparts.

The `Value_intf` signature, shown in Listing 4, defines the required operations for `Int32` and `Int64`. The `Int32` module specifies the following four elements:

- **`type`** `t`: Is the primary type of the module, such as `int32` for concrete execution.
- **`val`** `add`: A function to add two elements.
- **`val`** `gt`: A function to perform a greater-than comparison between two elements.
- **`val`** `eqz`: A function to check if an element is zero.

The `Int64` module specifies the equivalent types and functions for 64-bit integers.

Thanks to this parametrization, we can now easily implement variations of our interpreter where the basic operations are somewhat different. We could for example crash on integer overflow, or change the base functions for some that report statistics on how they are called.

### 3.3 The Monadic Interpreter

However, to do symbolic execution, we need to not only abstract over the types of values but also over the evaluation strategy itself. To do so, we need to scale the interpreter into a monadic form. Monads in functional programming represent

```ocaml
module Interpreter (Value : Value_intf) =
struct
  (* Makes items in the namespace of the Value module
  directly available here. *)
  open Value
  type value = I32 of Int32.t | I64 of Int64.t

  let select x =
    match x with I32 x -> not (Int32.eqz x) | _ -> false

  let eval_instr i stack =
    match i, stack with
    | Binop Add, I32 x :: I32 y :: stack ->
      I32 (Int32.add x y) :: stack
    | Binop Gt, I32 x :: I32 y :: stack ->
      I32 (Int32.gt x y) :: stack
    | If (if_true, if_false), cond :: stack ->
      let cond = select cond in
      if cond then eval_expr if_top stack
              else eval_expr if_bot stack
end
```

◼ **Listing 3** Parametric evaluator for the simplified Wasm syntax.

```ocaml
module type Value_intf = sig
  module Int32 : sig
    type t
    val add : t -> t -> t
    (* greater than *)
    val gt : t -> t -> t
    (* is equal to zero *)
    val eqz : t -> t
  end
  module Int64 : sig (* Same as Int32 *) end
end
```

◼ **Listing 4** Signature of the value module passed to the interpreter functor.

computations, and using them allows us to seamlessly support different execution strategies, including symbolic execution.

The monadic interpreter, shown in Listing 5, is parameterized over a module implementing the following monadic interface:

```ocaml
module type Choice_intf = sig
  type 'a t
  val return : 'a -> 'a t
```

```
4    val bind : 'a t -> ('a -> 'b t) -> 'b t
5    val select : Value.Int32.t -> bool t
6  end
```

Here, **type** `'a t` represents the monad type, where `t` is a type that takes a type parameter `'a`, allowing `t` to be used with various types (e.g., **int**, **bool**, etc.). The core functions in this signature are:

- `return : 'a -> 'a t`, which is the monadic *return* function. It takes a "bare" value and turns it into the "simplest" corresponding monadic value (cf. Figure 2);
- `bind : 'a t -> ('a -> 'b t) -> 'b t`, which is the monadic *bind* function. A function `f` going from bare values of type `'a` to monadic values `b M.t` can be seen as deriving a monadic meaning from the `'a` value and returning it on top of a bare value `'b`. *Bind* allows us to apply such an `f` function to an `'a` value already contained in the monad, combining the "monadic meaning" of both the input value and the one returned by `f` (cf. Figure 2);
- **val** `select : Value.Int32.t -> bool t`, which takes an abstract `Int32.t` and returns a monadic value containing a boolean. As before the returned value indicates whether the passed integer was non-zero. The fact that this value is monadic means that it represents a computation, and because it wraps a boolean value, it means that this computation "boils down" to a boolean. Select returning a monadic value is testimony to the fact that when reaching a branching point and having to decide on the truthness of an expression made of symbols, the symbolic execution can decide which cases to explore, and in which order.

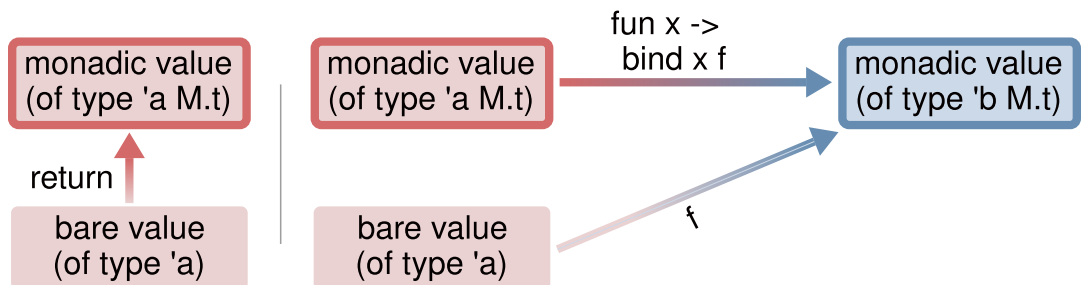

**Figure 2** A graphical representation of the two constitutive operations of a monad, *bind* and *return*. Monadic values are represented with a border, symbolizing how the monad "wraps" values. *Returns* wraps a non monadic value in the monad. *Bind* takes a function `f` that goes from bare values of type `'a` to monadic values containing `'b` (hence computing from the bare `'a` value both the `'b` value and a "monadic meaning") and turns it into a function going from monadic `a M.t` to monadic `b M.t` (hence combining the monadic meaning of `'a M.t` and the one derived by f).

The monadic evaluator differs from the parametric evaluator in two key ways:

1. **Monadic binding**: the custom **let*** operator wraps the monadic `bind`, enabling concise chaining of computations. This is similar to Haskell `do` notation, or Scala's `for` notation.

```
1  module Interpreter (Value : Value_intf) (Choice : Choice_intf) =
2  struct
3    (* Makes items in the namespace of the Value module
4    directly available here. *)
5    open Value
6    (* This is a syntax trick that allows to write: `let* x = e1 in e2`
7       Which will be desugared into: `Choice.bind e1 (fun x -> e2)`
8    *)
9    let ( let* ) v f = Choice.bind v f
10
11   let rec eval_instr i stack =
12     match i, stack with
13     | Binop Add, I32 x :: I32 y :: stack ->
14       Choice.return (I32 (Int32.add x y) :: stack)
15     | Binop Gt, I32 x :: I32 y :: stack ->
16       Choice.return (I32 (Int32.gt x y) :: stack)
17     | If (if_true, if_false), cond :: stack ->
18       let* cond = Choice.select cond in
19       if cond then eval if_true stack
20               else eval if_false stack
21 end
```

■ **Listing 5** Monadic evaluator for the simplified Wasm syntax.

2. Conditional evaluation: the condition is evaluated in a monadic context, supporting symbolic execution by deferring the evaluation strategy to the select function, which will select which branch to execute. This function returns a monadic computation of type **bool** Choice.t. Using **let***, we bind this result to a variable and then use it to choose between executing if_true or if_false.

For concrete execution, the monad is the identity monad, where computations are simply values:

```
1  type 'a t = 'a
2  let return x = x [@@inline]
3  let bind x f = f x [@@inline]
4  let select b =
5    match b with I32 x when x <> 0l -> true | _ -> false [@@inline]
```

Here, the code behaves exactly the same as it would without any abstraction, and there is no runtime cost thanks to the [@@inline] annotations.[3]

For symbolic execution, the monad involves additional complexity, as it needs to handle symbolic reasoning, combining features from both continuation and state monads. The detailed implementation of the symbolic choice monad will be covered in the next section.

---

[3] [@@inline] annotations are hints to the compiler that this function should be inlined. Inlining these small functions guarantees that the compiler will be able to optimize them further, making this identity monad a zero-cost abstraction.

In summary, the monadic functorization allows us to share most of the code between concrete and symbolic interpreters, enabling flexible evaluation strategies with minimal overhead. We believe this approach offers a convenient way to derive a symbolic interpreter from a concrete one in other contexts.

## 4 The Choice Monad

As detailed in Section 2.2, the symbolic interpreter frequently encounters branching choices, such as determining whether the condition of an if-then-else construct is `true` or `false`. Each choice corresponds to a different execution trace, and these traces form a prefix tree where each node represents a decision. Bugs, such as Wasm traps or assertion failures, reside within the leaves of this tree. While proving the complete absence of bugs would require exploring the entire tree — a generally infeasible task —, an efficient exploration of key branches can help identify bugs quickly.

To achieve this, we can aim for two goals: (a) maximize the exploration rate, *i.e.*, how many traces are evaluated per unit of time, and (b) use heuristics to prioritize the exploration of branches more likely to contain bugs. While our current implementation supports the foundational aspects needed for prioritization, defining the heuristics themselves is left for future research. In this section, we explain how our Choice monad ensures efficient branch exploration by leveraging parallel execution.

We first describe the structure of the Choice monad and its role in symbolic execution (Section 4.1). Then, we discuss our lazy memory model, which mitigates memory overconsumption issues that arise when scaling across multiple CPU cores (Section 4.2).

### 4.1 A Multicore Choice Monad

Our symbolic and parallel Choice monad exposes the same interface as described in the previous section, but integrates several key functionalities (each correponding to a monad transformer as illustrated in Figure 3):

- A **forkable cooperating coroutine monad**, where the execution can *yield* control to the scheduler and fork itself. By forking itself, the coroutine duplicates the execution and provides the notion of choice. Additionally, these coroutines have access to a worker-local storage (WLS), a value uniquely accessible to the worker executing the coroutine, which allows having "global" values of non thread-safe types.
- A **state monad**, which manages the interpreter's internal Wasm state and symbolic execution bookkeeping, particularly the path condition.
- An **error monad** that handles traps from Wasm programs and symbolic assertions failures. These errors propagate through the monadic structure.

The state and error monads closely resemble existing constructs in the literature [34], and we won't further explain them here. The coroutine monad has four base constructs, illustrated in Figure 4:

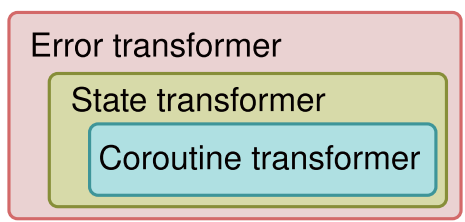

**Figure 3** Our choice monad is composed of three transformers layered together.

- *Choice* points indicate that the computation can take two directions, with different results.
- *Yield* points allow the scheduler to suspend the coroutine and start executing another one.
- *Value* leaves indicate a final value for this (sub-)coroutine. In our case, this value will be a value of the error monad.
- *Stop* leaves indicate that the execution stopped without a Value because this branch is actually unreachable.

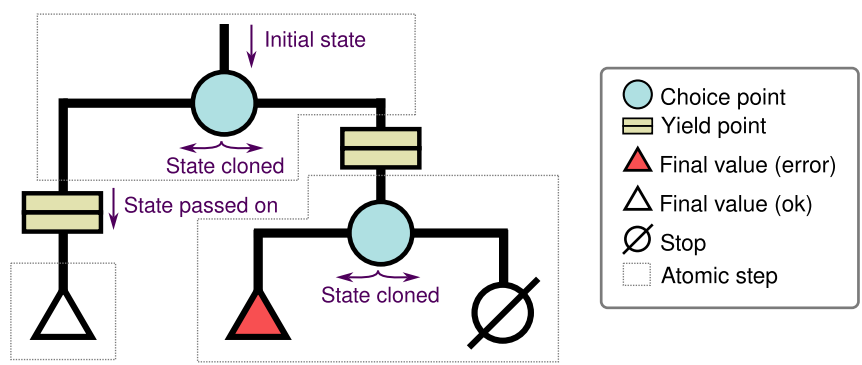

**Figure 4** A graphical representation of a value from our Choice monad

The monadic operations exposed by our Choice monad build a value representing the coroutine tree. This value describing a coroutine can then be passed to a scheduler which will run it in atomic steps bounded by yield points. When reaching a Value leaf, our scheduler jumps to a predetermined callback to handle this value properly.

#### 4.1.1 Implementation of the Multicore Choice Monad

The rest of this section is devoted to an in-depth presentation of our coroutine monad. Correctly implementing this monad was one of the technical challenges that once overcome allowed Owi to achieve the performances presented in Section 6, however understanding it exhaustively is not required to follow the rest of this article, so the less interested reader should feel free to skip directly to Section 4.2 (page 17).

**The Coroutine Monad**   The coroutine monad is implemented as follows in OCaml:

```ocaml
(* Two mutually recursive datatypes *)

(* The type of coroutines themselves,
   it has only one constructor, containing a function that
   takes the worker local storage and returns a status *)
type ('a, 'wls) t =
  Schedulable of ('wls -> ('a, 'wls) status)

(* The type of one step of the coroutine,
   it has four constructors *)
and ('a, 'wls) status =
  | Now of 'a (* Final value of type 'a *)
  | Yield of Prio.t * ('a, 'wls) t (* Coroutine with a priority *)
  (* Choice point *)
  | Choice of (('a, 'wls) status * ('a, 'wls) status)
  | Stop (* End of execution *)
```

Coroutines, of type t, execute in steps. Each (sub-)coroutine value (of type t) reads a worker local storage (WLS) and may result in:

- Now, indicating a final value.
- Yield, which schedules a continuation of the coroutine with an associated priority.
- Choice, denoting a choice between two execution paths,
- Stop, signaling the termination of the coroutine without any value.

The base primitives of this monad are straightforward, and are listed in Appendix A.1, to the exception of bind which is more complex and is provided below.

```ocaml
let rec bind (mx : ('a, 'wls) t) (f : 'a -> ('b, 'wls) t)
  : ('b, 'wls) t
  = Schedulable
    (fun wls ->
      (* A closure whose purpose is to traverse nested statuses and
         return the final value of one step of (bind mx f) *)
      let rec unfold_status (x : ('a, 'wls) status)
      : ('b, 'wls) status
        = match x with
        (* The final value is that of f applied to x *)
        | Now x -> run (f x) wls
        (* If the coroutine is yielding, we return a yield.
           The resulting next coroutine will be the result of
           recursively binding f to the initial next coroutine. *)
        | Yield (prio, lx) -> Yield (prio, bind lx f)
        (* On a choice, we simply unfold both branches. *)
        | Choice (mx1, mx2) ->
          let mx1' = unfold_status mx1 in
          let mx2' = unfold_status mx2 in
          Choice (mx1', mx2')
        (* We stop if mx stops *)
        | Stop -> Stop
```

```ocaml
23        in
24        unfold_status (run mx wls) )
```

**Scheduling Coroutines**    To execute coroutines, we use a scheduler that manages work distribution. This scheduler uses a synchronized FIFO queue to store coroutines and distributes work to available worker threads. The FIFO queue interface is provided in Appendix A.2. The scheduler itself is implemented as follows:

```ocaml
1  (* Our work queue type: a queue containing the Schedulable.t
2     coroutines defined above *)
3  type ('a, 'wls) work_queue = ('a, 'wls) Schedulable.t Wq.t
4
5  (* Scheduler type containing a work queue *)
6  type ('a, 'wls) t = { work_queue : ('a, 'wls) work_queue }
7
8  (* Create a new scheduler with an empty queue  *)
9  let make_scheduler () =
10   let work_queue = Wq.make () in
11   { work_queue }
12
13 (* Add a new task to the scheduler *)
14 let submit_task sched task = Wq.push task sched.work_queue
15
16 (* Main loop of a worker thread. Initialized with its WLS,
17    it runs steps of the coroutines scheduled on sched and calls
18    callback on their final value. *)
19 let work wls sched callback =
20   (* Handles the coroutine's status. The second argument is
21      the Queue callback allowing to push new elements. *)
22   let rec handle_status (t : _ Schedulable.status) write_back =
23     match t with
24     (* No more execution needed, return () *)
25     | Stop -> ()
26     (* A final value, pass it to the call back *)
27     | Now x -> callback x
28     (* The coroutine yielded, push the follow-up coroutine to
29        the queue. For now the priority is ignored. *)
30     | Yield (_prio, f) -> write_back f
31     (* The coroutine forked.
32        Evaluate each of the two sub coroutines sequentially. *)
33     | Choice (m1, m2) ->
34       handle_status m1 write_back;
35       handle_status m2 write_back
36   in
37   (* Use the queue work_while function to run each coroutine and
      ↪ handle their status while the scheduler queue is not empty *)
38   Wq.work_while
39     (fun f write_back ->
40       handle_status (Schedulable.run f wls) write_back)
```

```
41     sched.work_queue
42
43 (* Spawn a worker thread for scheduler `sched`. It will call callback
44    on each final value. callback_init and callback_close are the
45    callback initializer and deinitializer respectively. *)
46 let spawn_worker
47   sched wls_init callback callback_init callback_close =
48   callback_init ();
49   (* Start a Domain, an OCaml concept similar to a thread *)
50   Domain.spawn (fun () ->
51      Fun.protect
52        (* This will be executed even if the other closure
53           raises an exception *)
54        ~finally:callback_close
55        (* This closure is the actual thread code *)
56        (fun () ->
57           (* Initialize the worker local storage *)
58           let wls = wls_init () in
59           (* Run the worker loop defined above *)
60           try work wls sched callback
61           with e ->
62             let bt = Printexc.get_raw_backtrace () in
63             (* If this worker loop fails, mark the queue
64                as closed so that the scheduler stops *)
65             Wq.close sched.work_queue;
66             Printexc.raise_with_backtrace e bt ) )
```

The scheduler continually pulls coroutines from the work queue, runs them, and either finalizes their execution or reschedules them based on their current status (as illustrated in Appendix A.3). We currently treat all yields with equal priority, though supporting prioritized scheduling is a straightforward extension for future work.[4] Figuring what priority to give each branch is the tricky part and is left to future research.

**The Complete Choice Monad**  By applying state and error monad transformers, and fixing some of the type parameters, we get the final monadic type:

```
1 type 'a t = St of (Thread.t ->
2   Schedulable of ( Solver.t ->
3     ('a eval * Thread.t, Solver.t) status
4   )
5 )
```

Where:

- `Thread.t` represents the Wasm interpreter's state and the patch condition;

---

[4] To schedule with priority, we would simply change our queue module: for now it is backed by a FIFO queue, but we could change that to a priority queue while keeping the same multi-thread synchronization code.

- `Solver.t` represents the state of the SMT solver;
- `eval` represents the possibly failing result of an evaluation, and is defined as:

```
type 'a eval =
  | EVal of 'a (* success *)
  | ETrap of (* metadata describing the Wasm trap *)
  | EAssert of (* metadata representing the assertion failure *)
```

- `status` is the coroutine status defined above.

We lift all the required operations through the transformers, allowing us to define complex symbolic execution routines, `check` that takes a boolean value and uses the SMT solver to check its feasibility given the current assumptions, and `assume` that takes a boolean value and record that it was assumed to be true. Using those, we can now write a symbolic `select`:

```
(** Checks if a boolean value is feasible; if so, records it in the
    assumptions, otherwise halts the current branch. *)
let check_and_record (b : Symbolic_value.bool) =
  let* () = yield in
  if check b then
    record b
  else
    stop

let select (v : Symbolic_value.bool) : bool Choice.t =
  (* We first create two monadic values, each corresponding to a
     possible value of v (either true or false). Each of these
     branches simply checks the reachability, records the chosen
     value, and returns it as part of the monad. *)
  let true_branch =
    (* The let+ notation is similar to Haskell's do notation.
       This can be read:
         bind (check_and_record v) (fun () -> return true) *)
    let+ () = check_and_record v
    true
  in
  let false_branch =
    let+ () = check_and_record (Symbolic_value.not v) in
    false
  in
  (* Finally, creates the coroutine by combining both branches *)
  choose true_branch false_branch
```

### 4.2 The Lazy Memory Model

Symbolic execution often suffers from significant memory overhead due to the path explosion problem [2, 10]. When running Owi on the Test-Comp benchmarks [5] with a 30-second timeout, we observed up to 156,052 state duplications. A naive approach that duplicates the Wasm memory on every branch (the "state cloned" on Figure 4) would require storing potentially massive amounts of data, with Wasm memory ranging from 64,KiB to 4,GiB in size. This situation makes state duplication infeasible without optimized memory management.

Owi implements a lazy memory model that resembles a copy-on-write strategy. Instead of duplicating memory on every branch, new states share memory with their predecessors and track modifications in a separate map. When a memory location is accessed, Owi checks if it has been modified; if so, the value is retrieved from the modification map. Otherwise, the value is fetched from the original memory. This approach recursively applies to any prior states, reducing memory usage significantly.

This strategy is inspired by KLEE [9], though with a key difference: while KLEE models memory as a collection of objects and must copy entire objects even for small changes. In contrast, Owi models memory as a large array of bytes, allowing it to copy only the modified bytes. This is more efficient in terms of memory usage.

## 5 Demonstrating Owi's Symbolic Execution Capabilities

Having detailed how Owi is implemented, we now demonstrate its capabilities as a symbolic execution tool, first on Wasm and then on other languages.

### 5.1 Symbolic Execution of Wasm

We first illustrate Owi's ability by performing symbolic execution on the `$test_swap` function defined in Listing 1. To do so, we write the following code which calls the `$test_swap` function with two symbolic parameters. Remember that the `$test_swap` function already tests its own coherence using an `if` that leads to an `unreachable`.

```
(func $main
  call $i32_symbol ;; Creates a fresh symbol
  call $i32_symbol ;; Creates a fresh symbol
  call $test_swap
)
```

Calling Owi on this code, we can indeed identify the problematic input that leads to overflow.

```
$ owi sym test_swap.wat
Trap: unreachable
Model:
  (model
    (symbol_0 (i32 2147483646))
    (symbol_1 (i32 -2147483647)))
Reached problem!
```

Beyond this toy example, Owi can also find bugs in more realistic codebases. For example, we took the implementation of a B-tree data structure developed by C.

Costa [18], and the symbolic test suite used for the evaluation in (Section 6). Next, we added a few additional assertions to check the coherence of the data structure after a few operations. Running Owi determines that all possible execution paths are bug-free. As shown below by Owi outputting "All OK":

```
1 $ owi sym tree_3o3u.wat
2 All OK
```

Next, we introduce a bug in a critical section of the code by changing a "greater than" comparison to a "less than" one, and Owi manages to find that some execution path now leads to an assertion failure:

```
1  $ owi sym tree_3o3u_buggy.wat
2  Assert failure: false
3  Model:
4    (model
5      (symbol_0 (i32 -71872612))
6      (symbol_1 (i32 -72221176))
7      (symbol_2 (i32 -1543220269))
8      (symbol_3 (i32 -72326636))
9      (symbol_4 (i32 -205803988))
10     (symbol_5 (i32 71802348)))
11 Reached problem!
```

### 5.2 Symbolic Execution of C and Rust Programs

We have demonstrated Owi's capability to identify bugs in Wasm codebases. However, many languages, including C and Rust, can be compiled to Wasm. By exposing Owi's API to these host languages, we can use Owi to detect bugs in any program that can be compiled to Wasm.

To analyze a program written in a language $L$ using Owi, one must:

- **Implement the $L$ Primitives**: This involves modeling the program's interactions with its environment (such as disk and network I/O or system calls) for accurate analysis. This also includes correctly modeling dynamic memory management (e.g., `malloc`, `realloc`, and `free` in C).
- **Bind Owi Primitives**: This step involves connecting Owi's core primitives to $L$, enabling the generation of symbolic values, assertion checks, and controlled exploration.

Once these steps are completed, the program written in $L$ can be compiled to Wasm and executed in Owi for bug detection. However, it is crucial to note that some bugs, particularly those related to undefined behavior, might be obscured by the compiler. A Wasm program represents just one possible interpretation of the original program, which can have multiple interpretations if undefined behavior is present.

The following subsections outline our work in applying Owi to C and Rust.

### 5.2.1 Checking C Code

We have implemented portions of the C standard library, as well as a C header file that allows interaction with Owi from C. Additionally, we have added a subcommand to the Owi binary that facilitates the compilation of C programs to Wasm. With these components in place, C programs can be tested using Owi, as illustrated in the following examples.

A common approach to bug-finding is to write a test harness. For instance, to test a function f, we define symbolic inputs corresponding to its parameter types and perform assertions on its output. The harness is defined in the main function, with f being the function under test:

```c
#include <owi.h>

int f(int x, float y) {
  ...
}

void main(void) {
  int x = owi_i32();
  float y = owi_f32();

  int result = f(x, y);

  owi_assert(result == 42);
}
```

In this example, we define symbolic values for an integer x and a floating-point value y using Owi's exposed functions. We then apply f to these symbolic inputs and assert that the result is equal to 42. Owi's `owi_assert` function verifies that the symbolic expression provided as input always holds. Additionally, Owi offers an `owi_assume` function, which allows developers to restrict the path condition to certain scenarios, excluding uninteresting behaviors from exploration.

Alternatively, users can replace harnesses with *function contracts*, as seen in runtime assertion checking tools like E-ACSL [53] or deductive verification tools like Why3 [23]. These contracts are written in a specification logic distinct from the executable language, and translating them into symbolically executable code requires additional work. We have integrated such functionality for C by reusing E-ACSL. Our approach involves taking a C file annotated with E-ACSL specifications, generating an instrumented executable file via E-ACSL, and using a symbolic runtime to handle the instrumented code. Although this work is publicly available, we will not elaborate on it in this paper.

Owi can handle all Wasm features (except SIMD instructions) and thus complex C programs. For instance, the following C code example illustrates that Owi can handle function pointers in symbolic execution:

```c
#include <stddef.h>
#include <owi.h>

// fold(f, a, len, init) is
// f(f(... f(init, a[0]), a[1] ...), a[len-1])
int fold(int (*f)(int, int), int *array, size_t len, int init) {
  for (size_t i = 0; i < len; i++) {
    init = f(init, array[i]);
  }
  return init;
}

// a function whose result is expected
// to be positive if acc is positive
int step(int acc, int x) {
  return (acc + x * x);
}

void main(void) {
  // Create an array of unknown (symbolic) size
  // containing symbolic integers
  size_t len = owi_i32();
  int *array = malloc(sizeof(int) * len);
  for (size_t i = 0; i < len; i++) {
    array[i] = owi_i32();
  }

  // Apply the `fold` function on the symbolic array
  // with `step` as a function pointer
  int init = 42;
  int res = fold(step, array, len, init);

  // Check that the result is greater than the initial value
  // This should be the case because the output of `step` is expected
  //   to be positive when the accumulator is positive, which is the
  //   case of `init` here
  owi_assert(res >= init);

}
```

Although the code appears correct, Owi reveals a counterexample caused by integer overflow in step:

```
$ owi c function_pointer.c --fail-on-assertion-only -O0
Assert failure:
  (i32.ge (i32.add (i32 42) (i32.mul symbol_2 symbol_2)) (i32 42))
Model:
  (model
    (symbol_0 (i32 1))
```

```
7       (symbol_2 (i32 57344)))
8 Reached problem!
```

Running the code in unoptimized mode (-O0) ensures that the indirect call is not optimized away (Owi would still find the bug without it but there would not be a function reference in the Wasm code anymore). We use the `--fail-on-assertion-only` flag to focus the analysis on assertion failures rather than other potential bugs, such as exceeding Wasm's memory limits.

Similarly, we can use Owi to check the equivalence of two functions:

```c
#include <owi.h>

// check that f1 and f2 produce the same result for any inputs
void check_fn_equivalence(int (*f1)(int, int), int *(f2)(int, int)) {
  int x = owi_i32();
  int y = owi_i32();
  owi_assert(f1(x, y) == f2(x, y));
}

// two different implementations of the mean function
int mean1(int x, int y) {
  return (x & y) + ((x^y) >> 1);
}
int mean2(int x, int y) {
  return (x + y) / 2;
}

void main(void) {
  check_fn_equivalence(mean1, mean2);
}
```

Owi identifies that `mean1` and `mean2` are not equivalent due to integer overflow in `mean2`:

```
1 $ owi c ./function_equiv.c --fail-on-assertion-only -O0
2 Assert failure: (bool.eq (i32.add (i32.and symbol_0 symbol_1) (i32.shr
  ↪ (i32.xor symbol_0 symbol_1) (i32 1))) (i32.div (i32.add symbol_0
  ↪ symbol_1) (i32 2)))
3 Model:
4   (model
5     (symbol_0 (i32 -2147483648))
6     (symbol_1 (i32 -2147483646)))
7 Reached problem!
```

#### 5.2.2 Checking Equivalence of Mixed Rust and C code

Consider a scenario where a codebase is being migrated from C to Rust. In this codebase, the following C function computes the dot product of two 2D vectors:

```c
float dot_product(float x[2], float y[2]) {
    return (x[0]*y[0] + x[1]*y[1]);
}
```

A Rust developer might translate this into the following idiomatic Rust code using iterators:

```rust
fn dot_product_rust(x: &[f32; 2], y: &[f32; 2]) -> f32 {
    x.iter().zip(y).map(|(xi, yi)| xi * yi).sum()
}
```

To compare the behavior of the C and Rust implementations, we can bind the C function in Rust:

```rust
extern "C" {
    pub fn dot_product(x: *const f32, y: *const f32) -> f32;
}

fn dot_product_c(x: &[f32; 2], y: &[f32; 2]) -> f32 {
    unsafe { dot_product(x.as_ptr(), y.as_ptr()) }
}
```

Then, we use Owi to check that both implementations behave equivalently:

```rust
fn main() {
    let x = std::array::from_fn(|_| owi_sym::f32_symbol());
    let y = std::array::from_fn(|_| owi_sym::f32_symbol());
    let c_val = dot_product_c(&x, &y);
    let rust_val = dot_product_rust(&x, &y);
    owi_sym::assert(
      (c_val.is_nan() && rust_val.is_nan()) ||
      (c_val.to_bits() == rust_val.to_bits())
    )
}
```

Surprisingly, Owi finds a counterexample:

```
Model:
  (model
    (symbol_0 (f32 -0.))
    (symbol_1 (f32 -0.))
    (symbol_2 (f32 0.))
    (symbol_3 (f32 0.)))
```

This counterexample arises from the behavior of zeroes in IEEE 754 floating-point arithmetic. The following table illustrates the addition rules:

| +  | +0 | −0 |
|----|----|----|
| +0 | +0 | +0 |
| −0 | +0 | −0 |

The true neutral element of addition is −0, not +0. The Rust sum implementation behaves similarly to the following C code:

```c
1  float sum(float *x, int n) {
2      float sum = 0.0;
3      for (int i = 0; i < n; i++) {
4          sum += x[i];
5      }
6      return sum;
7  }
```

In this example, adding −0 to +0 leads to different results: Rust's sum yields +0, while C's implementation results in −0. Even though +0 and −0 compare equally, they can lead to significant differences in further computations. After discovering this issue, we identified it as a bug in Rust's standard library and submitted a pull request fixing it, which was accepted and merged.[5]

### 5.2.3 Asserting Properties on Rust Code Calling C Code.

Many foundational libraries are written in C, covering various usages: cryptography (NaCl and openssl), network (libcurl), graphics (GTK), common file formats (git, zlib, gzip), and so on. Newer programming languages often expose ways of binding to these libraries. However, doing so poses a challenge for verification because the codebase mixes several programming languages. Because Owi uses Wasm for verification it can handle code where one language (here Rust) calls another (here C) as long as both compile to Wasm.

To illustrate this, we implemented the following Rust code, calling a sha512 primitive from the popular libsodium library. In this code, the `user_entry` function takes a buffer, a user name, and a password and writes the user name, a null byte, and the sha512 hash of the password to the buffer.

```rust
1  extern "C" {
2      fn c_sha512(out: *mut u8, msg: *const u8, bytes: u64) -> c_int;
3  }
4
5  fn hash(out: &mut [u8], msg: &[u8]) {
6      let ret_code = unsafe {
7        c_sha512(out.as_mut_ptr(),msg.as_ptr(),msg.len() as u64)
8      };
9      if ret_code != 0 { panic!("Hash failed!") }
10 }
11
12 fn user_entry(mut output: &mut [u8], user_name: &str, pass: &str) {
13     use std::io::Write;
14     write!(&mut output, "{}\0", user_name.to_uppercase()).unwrap();
15     hash(output, pass.as_bytes());
16 }
```

To test it, we use the following main function. In this function we allocate a buffer big enough to store four bytes, the null byte, and the 64 bytes of the hash. We then

---

[5] https://github.com/rust-lang/rust/pull/129321 (visited on 2024-10-22).

generate a user name made of four characters, and call the `user_entry` function on those.

```
let user_name_size = 4;
let mut buffer = vec![0u8; user_name_size + 1 + 64];
let user_name =
    String::from_iter(std::iter::repeat_with(||
    ↪ owi_sym::char_symbol()).take(user_name_size));
user_entry(buffer.as_mut_slice(), &user_name, "password");
```

We have added no assertion to our main function, however, Owi still finds an issue:

```
Trap: memory heap buffer overflow
Model:
  (model
    (symbol_18 (i32 14))
    (symbol_19 (i32 8079))
    (symbol_20 (i32 0))
    (symbol_21 (i32 32)))
Reached problem!
```

Indeed, this example also demonstrates Owi's integrated memory tracking, and a buffer overflow has been correctly identified. That is because we allocate a buffer of 69 bytes, overlooking that the four Unicode characters (as given by `char_symbol`) may take up to four bytes each, so sixteen bytes in total. This buffer overflow cannot be caught by Rust itself (despite its safety properties) because it happens in the called C code: the user name and null byte will comfortably fit in the beginning of the 69 bytes long buffer but it is when writing the 64 bytes of the hash that we will overflow.

## 6 Experimental Evaluation

In this section, we evaluate the performance and effectiveness of Owi on Wasm and real-world C programs. Specifically, our evaluation aims to answer the two following research questions:

- **RQ1:** What is Owi's performance and how does it compare to other existing symbolic execution tools for Wasm?
- **RQ2:** How effective is Owi in detecting bugs and how does it compare to KLEE and Symbiotic?

### 6.1 Experimental Setup

To answer our research questions, we leverage two datasets of example programs. The first is a dataset of Wasm programs used for the performance evaluation of prior works [31, 40]. The second dataset, is a collection of real-world C programs used for evaluating software testing tools in the Test-Comp competition [5]. Next, we describe these datasets.

Owi: Performant Parallel Symbolic Execution Made Easy, an Application to WebAssembly- **Dataset 1 (B-tree test suite)**: Consists of 22 symbolic tests for a custom-made Wasm implementation of a B-tree data structure [18].
- **Dataset 2 (Test-Comp benchmarks)**: The 2024 edition of Test-Comp [5] consists of 11,042 tasks. The competition has two types of testing tasks: **(1)** *Cover-Branches* tasks, whose goal is to generate a set of concrete tests that covers the greatest possible number of program branches, and **(2)** *Cover-Error* tasks, whose goal is to generate at least one set of inputs that leads the execution of the program to a bug. Our focus is on the *Cover-Error* task, which includes 1,217 tasks out of the 11,042. However, we excluded two tasks due to one containing inline X86 assembly code and the other invalid code. Each task is a single C program.

To compare Owi with prior work, we set up the only three available stand-alone Wasm symbolic engines: WASP [40], SeeWasm [31], and Manticore [43]. Additionally, we evaluated the two best symbolic execution tools participating in the 2024 edition of Test-Comp: KLEE [9] and Symbiotic [14].

Our testbed consisted of an Ubuntu 22.04 server with an AMD EPYC 7451 24-Core Processor, offering 48 threads and 128GB of RAM. Owi was compiled with the OCaml 5.2.0 compiler using the Flambda1 optimizer. For the constraint solver, all compared tools used the Z3 SMT solver [20] version 4.13.0. The C code was compiled to Wasm using LLVM version 14 and using the flag -O3 in `clang`. For each execution of Owi, we used the -w24 flag, setting the number of workers to 24. The use of these flags and number of workers is experimentally justified in Appendix B.

The benchmarking code, reproducibility scripts, diagram generation, and cross-tool comparisons are available in Owi's GitHub repository.

**6.2 RQ1: Performance Evaluation**

To compare Owi's performance against Manticore, SeeWasm, and WASP, we use Dataset 1: the B-tree symbolic test suite from [40]. All tests share the same code template but vary in the number of symbolic values, with some constrained to be ordered. We denote the number of ordered and unordered symbolic values as $n_o$ and $n_u$, respectively.

**Results** Table 1 shows the speedup of Owi with 24 workers (`Owi-24`) compared to the tools identified in the prior work review: Manticore, SeeWasm, and WASP; as well as, Owi running with a single worker (`Owi-1`). The tests vary in both ordered symbolic values ($n_o$: from 2 to 9) and unordered symbolic values ($n_u$: from 1 to 3). For each ($n_o, n_u$) combination, the table provides the speedup of `Owi-24` against the raw execution time of each tool shown in Table 3. The speedup is claculated as $S = \frac{T_{tool}}{T_{\text{Owi}-24}}$, meaning that if $S$ is greater than 1, Owi was $S\times$ faster than *tool*.

Notably, as shown in Table 1, `Owi-24` consistently outperforms the other tools:

- Compared to Manticore, Owi achieves a speedup ranging from 17.2× to 844.3×, with an average of 312.6×.
- Compared to SeeWasm, Owi achieves a speedup ranging from 2.5× to 101.6×, with an average of 57.1×.

3:26

■ **Table 1** Speedup table calculated as $S_{tool} = \frac{T_{tool}}{T_{\texttt{Owi-24}}}$. Comparing Owi with 24 workers (`Owi-24`) against prior work tools: Manticore, SeeWasm, WASP, and single worker Owi (`Owi-1`). Each entry indicates the factor by which `Owi-24` was faster than the executor indicated at the top of the column.

| $n_o$ | $n_u = 1$ | | | | | $n_u = 2$ | | | | | $n_u = 3$ | | | | |
|---|---|---|---|---|---|---|---|---|---|---|---|---|---|---|---|
| | $S_{\texttt{Owi-24}}$ | $S_{\texttt{Mcore}}$ | $S_{\texttt{SW}}$ | $S_{\texttt{WASP}}$ | $S_{\texttt{Owi-1}}$ | $S_{\texttt{Owi-24}}$ | $S_{\texttt{Mcore}}$ | $S_{\texttt{SW}}$ | $S_{\texttt{WASP}}$ | $S_{\texttt{Owi-1}}$ | $S_{\texttt{Owi-24}}$ | $S_{\texttt{Mcore}}$ | $S_{\texttt{SW}}$ | $S_{\texttt{WASP}}$ | $S_{\texttt{Owi-1}}$ |
| 2 | 1.0 | 17.2 | 2.5 | 0.4 | 0.6 | 1.0 | 122.2 | 15.8 | 1.4 | 1.5 | 1.0 | 635.7 | 64.5 | 5.0 | 5.7 |
| 3 | 1.0 | 33.1 | 5.3 | 0.5 | 0.6 | 1.0 | 281.3 | 33.2 | 2.4 | 2.9 | 1.0 | 842.7 | 89.7 | 7.8 | 8.9 |
| 4 | 1.0 | 48.3 | 10.0 | 0.9 | 0.9 | 1.0 | 444.1 | 52.6 | 3.7 | 4.9 | 1.0 | 844.3 | 101.6 | 9.1 | 10.8 |
| 5 | 1.0 | 75.7 | 13.4 | 1.4 | 1.6 | 1.0 | 589.2 | 69.0 | 5.0 | 7.2 | 1.0 | 772.3 | 92.7 | 10.5 | 12.2 |
| 6 | 1.0 | 134.6 | 20.3 | 1.7 | 2.2 | 1.0 | 647.0 | 88.1 | 5.9 | 9.1 | 1.0 | 688.8 | 98.8 | 12.3 | 13.1 |
| 7 | 1.0 | 170.5 | 35.3 | 2.5 | 3.4 | 1.0 | 626.7 | 96.9 | 6.1 | 9.8 | 1.0 | 629.6 | 94.0 | 16.6 | 14.0 |
| 8 | 1.0 | 201.0 | 40.5 | 2.9 | 4.1 | 1.0 | 597.1 | 95.3 | 6.9 | 10.7 | – | – | – | – | – |
| 9 | 1.0 | 212.2 | 47.2 | 3.4 | 5.1 | 1.0 | 564.0 | 89.1 | 6.8 | 11.2 | – | – | – | – | – |

- Compared to WASP, Owi achieves a speedup ranging from 0.4× to 16.4×, with an average of 4.1×.
- Compared to Owi with one worker, Owi achieves a speedup ranging from 0.6× to 14.0×, with an average of 4.5×.

Furthermore, Table 1 reveals three minor observations:

1. For $2 \leq n_o \leq 4$ and $n_u = 1$, `Owi-24` is slower than both WASP and `Owi-1`. This slowdown is caused by the overhead of creating and destroying workers, and the synchronization between them.
2. In general, the speedup achieved by `Owi-24` is greater against `Owi-1` than against WASP. This is because Owi is not optimized for single-worker performance, resulting in occasional slower performance in single-worker mode compared to WASP, hinting at potential room for improvement.
3. Speedup improves with an increasing number of symbolic variables due to the exponential growth in explored paths. Benefiting the parallel exploration of Owi.

### 6.3 RQ2: Effectiveness in Bug Detection

We assess Owi's effectiveness in detecting bugs and compare it to KLEE and Symbiotic by running all tools on Dataset 2: Test-Comp's Cover-Error tasks. While Test-Comp has a 900-second limit per task, we use a 30-second limit to keep total runtimes under 8 hours (as these tasks are very memory intensive). This 30-second limit was still enough for both KLEE and Symbiotic to solve over 80% of the tasks in the original competition.[6] The results are summarized in Table 2.

---

[6] https://test-comp.sosy-lab.org/2024/results/results-verified/META_Cover-Error.table.html#/quantile?selection=cpu (visited on 2024-10-22).

- **Table 2** Solved tasks by KLEE, Owi, and Symbiotic on the *Cover-Error* Test-Comp benchmarks.

| Tool | Reached | Timeout | Nothing | $\frac{\text{Nothing}}{\text{Reached+Nothing}}$ |
| --- | --- | --- | --- | --- |
| KLEE | 782 | 368 | 65 | 7.67 % |
| Owi | 676 | 539 | 0 | 0.00 % |
| Symbiotic | 489 | 657 | 69 | 12.37 % |

**Reached Bugs**   A bug is considered "Reached" when a tool generates a concrete set of inputs that causes the program to reach a location marked with a failure, such as an **assert(**false**)**.

The "Reached" column in Table 2 shows that KLEE detects the most bugs, while Owi finds 0.86× as many. KLEE benefits from various exploration strategies, whereas Owi currently lacks efficient exploration heuristics and explores paths as they are discovered. This strategic difference has a significant impact on performance in symbolic execution engines, which likely explains the detection gap between the two tools. Moreover, KLEE is using the STP SMT solver and not Z3, since they noticed it was leading to much better results for symbolic execution. Trying the STP solver in Owi is left to future work.

Despite this, Owi outperforms Symbiotic, detecting 1.38× more bugs. Considering that KLEE and Symbiotic ranked $2^{\text{nd}}$ and $3^{\text{rd}}$,[7] respectively, among the 20 tools tested in the 2024 Test-Comp, Owi's performance demonstrates that its bug-finding capabilities are competitive with top-tier software testing tools.

**Undetected Bugs**   A bug is considered undetected when a tool terminates without timeout and without producing any bug-causing input, even if a bug is present. This is referred to as a *false negative*.

The "Nothing" column in Table 2 shows the number of false negatives reported by each tool, while the last column shows the percentage of false negatives relative to all non-timeout answers (i.e. Reached + Nothing). Notably, KLEE has a false negative rate of 7.67 %, and Symbiotic has a rate of 12.37 %. In contrast, Owi produces no false negatives.

This is because KLEE deliberately under-approximates certain C standard library functions. For example, when a program allocates memory with a symbolic size using `malloc`, KLEE provides only one concrete memory chunk [7]. This approach can accelerate execution and increase the number of bugs detected within the given timeout—provided the bug is independent of the chunk's size—but it also introduces potential false negatives. Since Symbiotic uses KLEE for symbolic execution, it will also suffer from the same potential false negatives.

---

[7] Two variants of FuseBMC ranked first.

## 7 Related Work

The aim of this work is to develop a scalable and maintainable Wasm interpreter capable of performing efficient symbolic execution. Our contributions are closely aligned with research on monadic and parametric interpreters as well as symbolic execution tools specifically designed for Wasm.

**Monadic and Parametric Interpreters** Our approach shares similarities with the work by Mensing et al. [42], who derive a symbolic execution engine from a definitional interpreter. Like our Wasm interpreter, their approach begins with a concrete implementation and evolves into a parametric one with an abstract value interface. They instantiate the interpreter in both concrete and symbolic modes. The primary distinction between our work and theirs is the focus on dynamically-typed languages with recursive functions and pattern matching. Additionally, they adopt a breadth-first search strategy in their symbolic interpreter to explore the program's state space. In contrast, our implementation leverages a co-routine monad, which allows us to prioritize the exploration of more pertinent branches of execution.

Necro [45], a framework for formalizing programming-language semantics, generates interpreters or proofs in Coq [54] based on language semantics. Necro generates an OCaml functor that is parametric over an interpretation monad handling applications and branches. Like our approach, it leverages monads to manage computational effects, though Necro focuses on concrete interpretation rather than extending to symbolic execution as we do.

**Symbolic Execution for Wasm** Symbolic execution has been extensively employed to uncover critical errors and vulnerabilities across various programming languages, including C [26], C++ [9], Java [50], and Python [15]. For the Web, several state-of-the-art tools for symbolic execution of JavaScript code [25, 38, 48, 49, 51] demonstrate the demand for such tools in validating and testing modern web applications.

Symbolic execution engines can generally be divided into two classes: static and dynamic/concolic execution engines [2]. Static symbolic execution tools [25, 35, 36, 46, 48, 55] explore the entire symbolic execution tree up to a predefined depth. On the other hand, concolic execution tools [9, 13, 26, 38, 49, 50, 51] combine concrete and symbolic executions, focusing on exploring one path at a time. Numerous symbolic execution tools have been proposed for various programming languages, as surveyed in [2, 10, 11]. Below, we delve into the current symbolic execution tools for Wasm, aside from Owi.

WANA [33] is a cross-platform tool that employs static symbolic execution to detect vulnerabilities in smart contracts compiled to Wasm bytecode. However, it lacks a stand-alone symbolic execution engine for general Wasm code. Manticore [43], a flexible symbolic execution framework for binaries and smart contracts, includes support for Wasm bytecode, but relies on complex, manually crafted Python scripts for each test. Manticore is no longer actively maintained. WASP [40], built on an old Wasm reference interpreter, uses concolic execution to reduce solver interactions and simplify memory modeling. However, it remains limited to Wasm 1.0, as updating the interpreter to support newer language features was deemed infeasible by the development team. SeeWasm [31] introduces in [32] a novel symbolic execution approach with fine-grained local search strategies, but the need for deep program-specific insights make its practical applicability limited.

None of these tools, including WANA, Manticore, WASP, and SeeWasm, support parallel or concurrent state space exploration for Wasm programs. Owi, however, is the first symbolic execution engine for Wasm to introduce parallel state space exploration, utilizing the power of OCaml-Multicore.

**Verification of Cross-language Codebases**   Verification of cross-language codebases requires that both languages have a common compilation target supported by the verification tool. Hardware assembly languages are notoriously hard to run verification on, especially because control flow is often intricate and hard to reconstruct in the assembly code generated by compilers. LLVM IR is the target of some verification tools (e.g. KLEE), which could be used to do verification of cross-language code, but we couldn't find examples of it in the literature. A possible explanation to this is the lack of tooling to easily generate and manipulate LLVM bitcode. On the other hand, Wasm is supported by an ever-increasing number of languages, and its design lends itself well to verification and symbolic execution. Moreover, the languages's standard libraries and runtime are already distributed as Wasm, which is not the case of LLVM IR.

**Parallel Symbolic Execution Engines**   Extensive literature exists on parallelizing symbolic execution [21, 52, 56]. The articles we found all use a similar algorithm to ours where sub-branches of the execution tree are dispatched to different worker threads, but none of them seemed to have our lazy memory model allowing efficient scaling, and as these tools did not participate in TestComp, there is no easy way to run them on our benchmark set, which would require to modify these tools so that they provide a C interface compatible with the one expected by TestComp files. Moreover, to the best of our knowledge, there is no tool that is both well advertised and usable for Wasm, C and/or Rust, and able to work in parallel. Finally, none of the tools we found are using a monad based software architecture.

## 8   Conclusion and Future Directions

With Owi now capable of parallel symbolic execution, delivering promising results, several opportunities for enhancement can further improve its bug-finding capabilities. Our immediate priority is to finalize the implementation of a concolic monad for Owi. Although not yet multi-core and still harboring a few bugs, the current version performs on par with the symbolic version when executed with a single worker.

Next, we plan to implement a more efficient exploration strategy that capitalizes on the prioritization enabled by our monadic design, potentially incorporating techniques such as A* [19]. To address the combinatorial explosion of execution paths — especially in loops — we intend to employ constrained Horn clauses (CHC) [28] for inferring loop invariants and hash-consed Patricia trees [22] to merge states more effectively.

Another key objective is integrating the WasmGC proposal into Owi, which will enable the symbolic execution of OCaml code (by compiling it with Wasocaml, our OCaml-to-WasmGC compiler), or other languages such as Java or Scala.

In conclusion, our work illustrates that Wasm is a robust target for symbolic execution, particularly given the increasing number of languages compiling to Wasm. The Wasm ecosystem allows us to leverage existing toolchains across languages, facilitating cross-language symbolic execution. Our approach also offers a reusable technique for constructing high-performance symbolic execution engines by extending a concrete interpreter into a monadic one with a streamlined choice monad implementation. This method is adaptable to other contexts and has already demonstrated strong performance in our evaluations. We are confident that incorporating more advanced symbolic execution techniques into Owi will lead to significant improvements and impactful results.

**Acknowledgements** First, we would like to thank the reviewers for their valuable comments. Moreover, we are grateful to Olivier Pierre, Boris Eng and Fabrice Le Fessant for their feedback on the article. This work was supported by national funds through Fundação para a Ciência e a Tecnologia (FCT) via grant PRT/BD/154029/2022.

## A  The Choice Monad

### A.1  Base Primitives

```
(* Convert a value to a coroutine that returns it immediately *)
let return (x : 'a) : ('a,'wls) t =
  (* A Schedulable value containing a function ignoring the WLS
  and returning x *)
  Schedulable (fun _wls -> Now x)

(* Executes one step of the coroutine with the provided WLS *)
let run (Schedulable mxf : ('a, 'wls) t) (wls : 'wls)
  : ('a, 'wls) status
  = mxf wls

(* Yields control back to the scheduler with a specified priority *)
let yield (prio: Prio.t) : (() ,'wls) t = u
  Schedulable (fun _wls -> Yield (prio, return ()))

(** Creates a choice between two coroutines *)
let choose (a : ('a, 'wls) t) (b : ('a, 'wls) t) : ('a, 'wls) t =
  Schedulable (fun wls -> Choice (run a wls, run b wls))

(* Accesses the worker-local storage *)
let wls : ('wls,'wls) t =
  (* This coroutine inner function simply returns the WLS *)
  Schedulable (fun wls -> Now wls)

(* Forks the current coroutine, yielding back control to the scheduler.
   The parent coroutine will resume execution with a priority
   prio_parent, and the child coroutine with a priority prio_child.
   In the parent coroutine, fork is false (resp. child is true) *)
let fork (prio_parent: Prio.t) (prio_child : Prio.t): bool t =
  Schedulable (fun _wls ->
       Choose
         (Yield (prio_parent, Now false))
         (Yield (prio_child, Now true)))
```

### A.2 The FIFO Queue Interface

```ocaml
(** Synchronized FIFO queues *)

(** The main and only type of this module. *)
type !'a t

(** Create a new queue *)
val make : unit -> 'a t

(** Add a new element to the queue *)
val push : 'a -> 'a t -> unit

(** Get an element from the queue.
    The boolean shall be true to atomically start a new pledge
    (cf make_pledge) while popping. *)
val pop : 'a t -> bool -> 'a option

(** Make a new pledge, i.e. indicate that elements may be pushed to
    the queue and that calls to pop should block waiting for them *)
val make_pledge : 'a t -> unit

(** End one pledge *)
val end_pledge : 'a t -> unit

(** Mark the queue as closed: all threads trying to pop from it will
    get no element *)
val close : 'a t -> unit

(** Call in a loop the provided function while there is elements
    in the queue. The provided function should take a queue element
    as first parameter, and a callback allowing to push new elements
    to the queue as second parameter. *)
val work_while : ('a -> ('a -> unit) -> unit) -> 'a t -> unit
```

## A.3 Illustration of the Scheduler

The set of atomic tasks to be scheduled

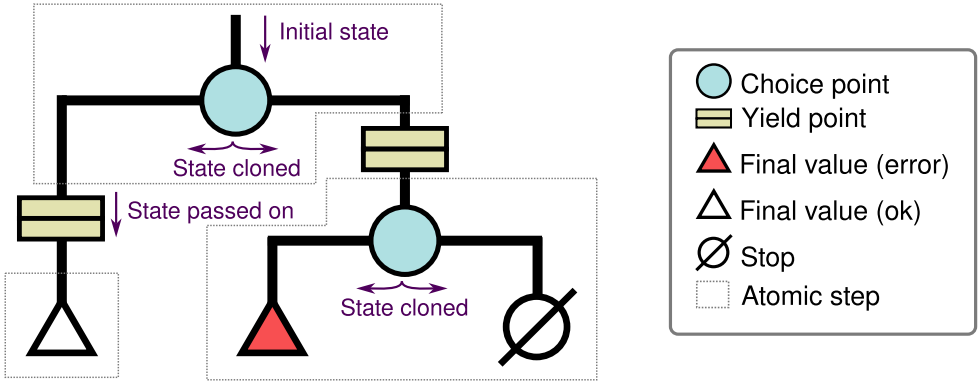

The initial state is put in the work queue

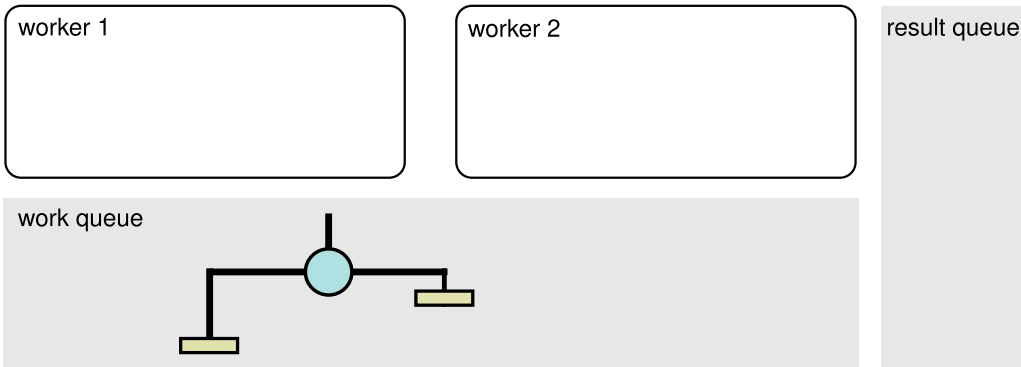

A worker takes this first item from the work queue

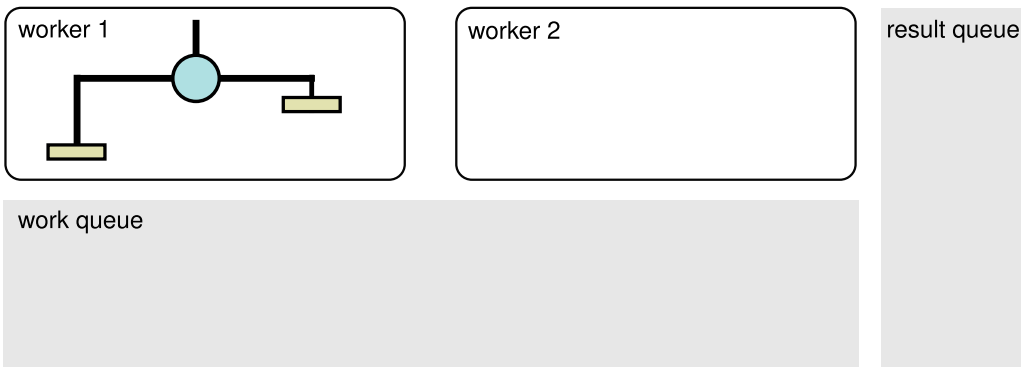

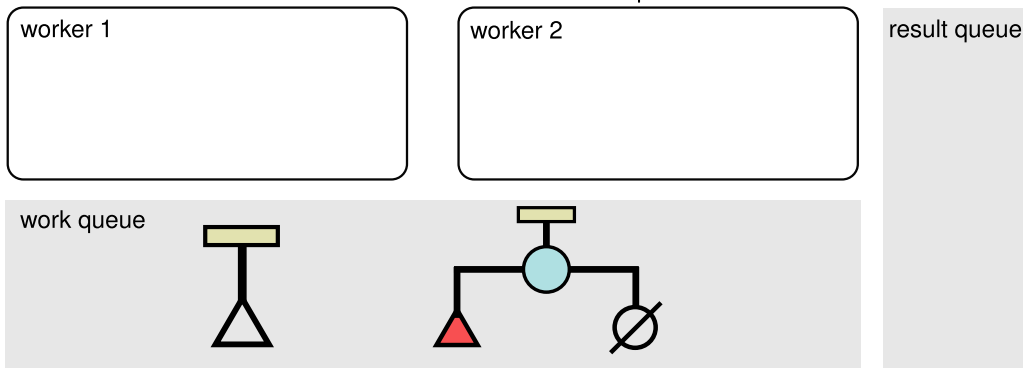

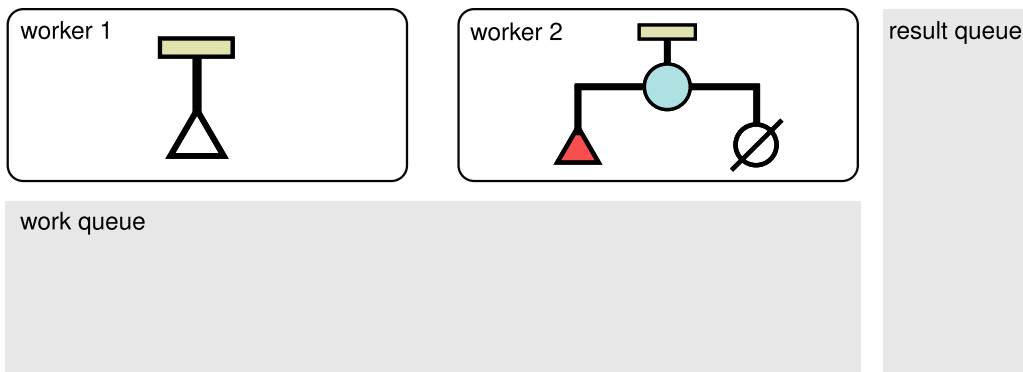

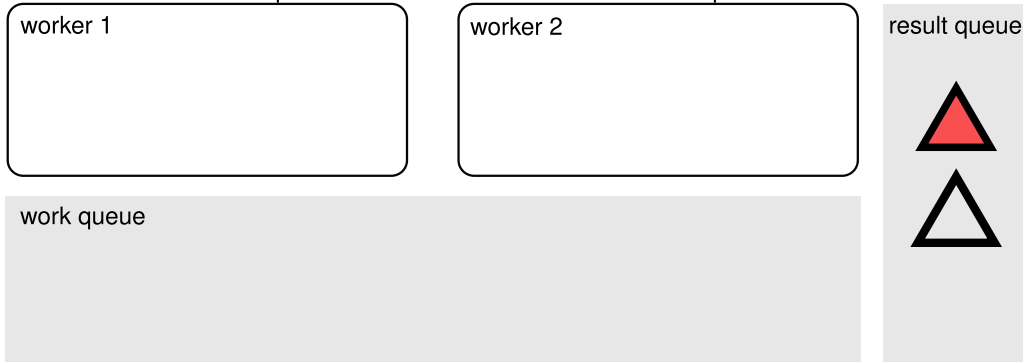

## B  Exploring Different Parameters

### B.1  Multicore Evaluation

To assess the efficiency of Owi's multicore implementation, we varied the number of workers from `-w1` to `-w48` and observed the results on the Test-Comp Cover-Error benchmarks, which are presented in the following table:

| Number of workers | 1 | 2 | 4 | 8 | 12 | 16 | 20 | **24** | 36 | 48 |
|---|---|---|---|---|---|---|---|---|---|---|
| Reached bugs | 612 | 645 | 663 | 664 | 675 | 675 | 675 | **676** | 675 | 672 |

The results indicate that as the number of workers increases, so does the number of tasks solved, peaking at 24 workers. Beyond this optimal point, performance slightly declines, likely due to increased cache contention of virtual threads sharing the same logical core. A comparison of runs using 1 and 24 workers shows that the multicore version solved 62 more tasks, representing a 10% improvement. The distribution of execution times remained consistent across all worker configurations.

### B.2 Compiler Optimization Levels

We also experimented with different `clang` optimization levels (`-O0` to `-O3`) to evaluate their impact on performance. The results on the Test-Comp benchmarks are summarized in the table below:

| Clang Optimisation level | O0 | O1 | O2 | **O3** |
|---|---|---|---|---|
| Reached bugs | 554 | 674 | 674 | **676** |

At the `-O0` optimization level, Owi solved fewer tasks. Most unsolved tasks in this category were due to loops that were not optimized, causing the exploration process to stall. Beginning at `-O1`, Owi solved a higher number of tasks, indicating that optimizations designed for concrete execution also benefit symbolic execution. Even with the increased compilation time, optimizations positively impacted the number of completed tasks.

## C  B-tree Raw Execution Times

**Table 3** Time results for Manticore ($T_{\text{Mcore}}$), SeeWasm ($T_{\text{SW}}$), WASP ($T_{\text{WASP}}$), single threaded Owi ($T_{\text{Owi}}$), and Owi with 24 workers ($T_{\text{Owi24}}$) on the B-tree test suite.

| $n_o$ | $n_u = 1$ | | | | | $n_u = 2$ | | | | | $n_u = 3$ | | | | |
|---|---|---|---|---|---|---|---|---|---|---|---|---|---|---|---|
| | $T_{\text{Mcore}}$ | $T_{\text{SW}}$ | $T_{\text{WASP}}$ | $T_{\text{Owi}}$ | $T_{\text{Owi24}}$ | $T_{\text{Mcore}}$ | $T_{\text{SW}}$ | $T_{\text{WASP}}$ | $T_{\text{Owi}}$ | $T_{\text{Owi24}}$ | $T_{\text{Mcore}}$ | $T_{\text{SW}}$ | $T_{\text{WASP}}$ | $T_{\text{Owi}}$ | $T_{\text{Owi24}}$ |
| 2 | 4.356 | 0.634 | 0.104 | 0.155 | 0.254 | 35.058 | 4.534 | 0.391 | 0.441 | 0.287 | 382.687 | 38.829 | 3.062 | 3.466 | 0.602 |
| 3 | 11.151 | 1.791 | 0.166 | 0.218 | 0.337 | 110.546 | 13.053 | 0.953 | 1.159 | 0.393 | 1,117.439 | 118.925 | 10.399 | 11.863 | 1.326 |
| 4 | 17.659 | 3.648 | 0.330 | 0.339 | 0.366 | 261.156 | 30.920 | 2.178 | 2.895 | 0.588 | 2,713.705 | 326.637 | 29.403 | 34.852 | 3.214 |
| 5 | 33.305 | 5.897 | 0.607 | 0.712 | 0.440 | 507.321 | 59.453 | 4.337 | 6.197 | 0.861 | 5,722.794 | 686.920 | 77.568 | 90.649 | 7.410 |
| 6 | 67.175 | 10.137 | 0.826 | 1.081 | 0.499 | 896.068 | 121.961 | 8.179 | 12.593 | 1.385 | 11,118.566 | 1,595.489 | 198.750 | 211.120 | 16.142 |
| 7 | 97.692 | 20.221 | 1.458 | 1.952 | 0.573 | 1,584.262 | 244.923 | 15.520 | 24.684 | 2.528 | 20,727.641 | 3,096.070 | 538.423 | 460.280 | 32.924 |
| 8 | 152.363 | 30.762 | 2.165 | 3.117 | 0.758 | 2,567.524 | 409.891 | 29.648 | 45.938 | 4.300 | – | – | – | – | – |
| 9 | 210.069 | 46.709 | 3.390 | 5.062 | 0.990 | 3871.196 | 611.926 | 46.851 | 76.744 | 6.865 | – | – | – | – | – |

**References**

[1] Syrus Akbary and Ivan Enderlin. *Wasmer: The Universal WebAssembly Runtime*. URL: https://wasmer.io/ (visited on 2024-10-24).

[2] Roberto Baldoni, Emilio Coppa, Daniele Cono D'elia, Camil Demetrescu, and Irene Finocchi. "A Survey of Symbolic Execution Techniques". In: *ACM Computing Surveys* (2018). DOI: 10.1145/3182657.

[3] Haniel Barbosa, Clark Barrett, Martin Brain, Gereon Kremer, Hanna Lachnitt, Makai Mann, Abdalrhman Mohamed, Mudathir Mohamed, Aina Niemetz, Andres Nötzli, Alex Ozdemir, Mathias Preiner, Andrew Reynolds, Ying Sheng, Cesare Tinelli, and Yoni Zohar. "cvc5: A Versatile and Industrial-Strength SMT Solver". In: *Tools and Algorithms for the Construction and Analysis of Systems*. TACAS. Springer International Publishing, 2022, pages 415–442. ISBN: 978-3-030-99524-9. DOI: 10.1007/978-3-030-99524-9_24.

[4] John Bergbom. *Memory safety: old vulnerabilities become new with WebAssembly*. Technical report. Forcepoint, 2018.

[5] Dirk Beyer. "Software Testing: 5th Comparative Evaluation: Test-Comp 2023". en. In: *Fundamental Approaches to Software Engineering*. FASE. Cham: Springer Nature Switzerland, 2023, pages 309–323. ISBN: 978-3-031-30826-0. DOI: 10.1007/978-3-031-30826-0_17.

[6] François Bobot, Bruno Marre, Guillaume Bury, Stéphane Graham-Lengrand, and Hichem Rami Ait El Hara. *Colibri2*. URL: https://colibri.frama-c.com (visited on 2024-10-22).

[7] Luca Borzacchiello, Emilio Coppa, Daniele Cono D'Elia, and Camil Demetrescu. "Memory models in symbolic execution: key ideas and new thoughts". In: *Software Testing, Verification and Reliability* (2019). DOI: 10.1002/stvr.1722.

[8] Ruben Bridgewate. *Node v12.3.0*. URL: https://nodejs.org/en/blog/release/v12.3.0 (visited on 2024-04-27).

[9] Cristian Cadar, Daniel Dunbar, and Dawson Engler. "KLEE: unassisted and automatic generation of high-coverage tests for complex systems programs". In: *USENIX Conference on Operating Systems Design and Implementation*. OSDI. USA: USENIX Association, 2008, pages 209–224. DOI: 10.5555/1855741.1855756.

[10] Cristian Cadar, Patrice Godefroid, Sarfraz Khurshid, Corina S. Păsăreanu, Koushik Sen, Nikolai Tillmann, and Willem Visser. "Symbolic execution for software testing in practice: preliminary assessment". In: *International Conference on Software Engineering*. ICSE. New York, NY, USA: Association for Computing Machinery, 2011, pages 1066–1071. ISBN: 978-1-4503-0445-0. DOI: 10.1145/1985793.1985995.

[11] Cristian Cadar and Koushik Sen. "Symbolic execution for software testing: three decades later". In: *Communications of the ACM* (2013), pages 82–90. DOI: 10.1145/2408776.2408795.

[12] Can I Use. *WebAssembly Browser Support*. URL: https://caniuse.com/wasm (visited on 2024-10-24).

[13] Sang Kil Cha, Thanassis Avgerinos, Alexandre Rebert, and David Brumley. "Unleashing Mayhem on Binary Code". In: *IEEE Symposium on Security and Privacy*. SP. 2012, pages 380–394. DOI: 10.1109/SP.2012.31.

[14] Marek Chalupa, Jakub Novák, and Jan Strejček. "Symbiotic 8: Parallel and Targeted Test Generation". In: *Fundamental Approaches to Software Engineering*. Place: Cham. Springer International Publishing, 2021, pages 368–372. ISBN: 978-3-030-71500-7. DOI: 10.1007/978-3-030-71500-7_20.

[15] Ting Chen, Xiao-song Zhang, Rui-dong Chen, Bo Yang, and Yang Bai. "Conpy: Concolic Execution Engine for Python Applications". In: *Algorithms and Architectures for Parallel Processing*. ICA3PP. Cham: Springer International Publishing, 2014, pages 150–163. ISBN: 978-3-319-11194-0. DOI: 10.1007/978-3-319-11194-0_12.

[16] Lin Clark. *Standardizing WASI: A system interface to run WebAssembly outside the web*. URL: https://hacks.mozilla.org/2019/03/standardizing-wasi-a-webassembly-system-interface (visited on 2024-10-24).

[17] Sylvain Conchon, Albin Coquereau, Mohamed Iguernlala, and Alain Mebsout. "Alt-Ergo 2.2". In: *SMT Workshop: International Workshop on Satisfiability Modulo Theories*. 2018.

[18] Carolina Costa. "Concolic Execution for WebAssembly". Master's thesis. Instituto Superior Técnico, 2020. URL: https://fenix.tecnico.ulisboa.pt/cursos/meic-a/dissertacao/846778572212567 (visited on 2024-10-22).

[19] Theo De Castro Pinto, Antoine Rollet, Grégoire Sutre, and Ireneusz Tobor. "Guiding Symbolic Execution with A-Star". In: *Software Engineering and Formal Methods*. SEFM. Berlin, Heidelberg: Springer-Verlag, 2023, pages 47–65. ISBN: 978-3-031-47114-8. DOI: 10.1007/978-3-031-47115-5_4.

[20] Leonardo De Moura and Nikolaj Bjørner. "Z3: An efficient SMT solver". In: *Tools and Algorithms for the Construction and Analysis of Systems*. TACAS. Springer, 2008, pages 337–340. ISBN: 978-3-540-78799-0. DOI: 0.1007/978-3-540-78800-3_24.

[21] Amir Mohammad Deilami. "KLEEWT: A parallel symbolic execution engine". Publisher: Simon Fraser University. Master's thesis. Simon Fraser University, 2023. URL: https://summit.sfu.ca/item/38004 (visited on 2024-10-22).

[22] Jean-Christophe Filliâtre and Sylvain Conchon. "Type-safe modular hash-consing". In: *Proceedings of the 2006 workshop on ML*. ML. New York, NY, USA: Association for Computing Machinery, 2006, pages 12–19. DOI: 10.1145/1159876.1159880.

[23] Jean-Christophe Filliâtre and Andrei Paskevich. "Why3 — Where Programs Meet Provers". In: *Programming Languages and Systems*. ESOP. Berlin, Heidelberg: Springer, 2013, pages 125–128. ISBN: 978-3-642-37036-6. DOI: 10.1007/978-3-642-37036-6_8.

[24] José Fragoso Santos, Petar Maksimović, Sacha-Élie Ayoun, and Philippa Gardner. "Gillian, part i: a multi-language platform for symbolic execution". In: *Programming Language Design and Implementation*. PLDI. New York, NY, USA: Association for Computing Machinery, 2020, pages 927–942. ISBN: 978-1-4503-7613-6. DOI: 10.1145/3385412.3386014.

[25] José Fragoso Santos, Petar Maksimović, Gabriela Sampaio, and Philippa Gardner. "JaVerT 2.0: compositional symbolic execution for JavaScript". In: *Proceedings of the ACM on Programming Languagues* POPL (2019). DOI: 10.1145/3290379.

[26] Patrice Godefroid, Nils Klarlund, and Koushik Sen. "DART: directed automated random testing". In: *Programming Language Design and Implementation*. PLDI. New York, NY, USA: Association for Computing Machinery, 2005, pages 213–223. ISBN: 1-59593-056-6. DOI: 10.1145/1065010.1065036.

[27] GoogleSecurityResearch. *Google Chrome 73.0.3683.103 - 'WasmMemoryObject::Grow' Use-After-Free*. URL: https://www.exploit-db.com/exploits/46968 (visited on 2024-10-24).

[28] Arie Gurfinkel and Nikolaj Bjørner. "The Science, Art, and Magic of Constrained Horn Clauses". In: *International Symposium on Symbolic and Numeric Algorithms for Scientific Computing*. SYNASC. 2019, pages 6–10. DOI: 10.1109/SYNASC49474.2019.00010.

[29] Andreas Haas, Andreas Rossberg, Derek L. Schuff, Ben L. Titzer, Michael Holman, Dan Gohman, Luke Wagner, Alon Zakai, and JF Bastien. "Bringing the web up to speed with WebAssembly". In: *Programming Language Design and Implementation*. PLDI. New York, NY, USA: Association for Computing Machinery, 2017, pages 185–200. ISBN: 978-1-4503-4988-8. DOI: 10.1145/3062341.3062363.

[30] Adam Hall and Umakishore Ramachandran. "An execution model for serverless functions at the edge". In: *International Conference on Internet of Things Design and Implementation*. IoTDI. New York, NY, USA: Association for Computing Machinery, 2019, pages 225–236. ISBN: 978-1-4503-6283-2. DOI: 10.1145/3302505.3310084.

[31] Ningyu He, Zhehao Zhao, Hanqin Guan, Jikai Wang, Shuo Peng, Ding Li, Haoyu Wang, Xiangqun Chen, and Yao Guo. "SeeWasm: An Efficient and Fully-Functional Symbolic Execution Engine for WebAssembly Binaries". In: *International Symposium on Software Testing and Analysis*. ISSTA. New York, NY, USA: Association for Computing Machinery, 2024, pages 1816–1820. ISBN: 9798400706127. DOI: 10.1145/3650212.3685300.

[32] Ningyu He, Zhehao Zhao, Jikai Wang, Yubin Hu, Shengjian Guo, Haoyu Wang, Guangtai Liang, Ding Li, Xiangqun Chen, and Yao Guo. "Eunomia: Enabling User-Specified Fine-Grained Search in Symbolically Executing WebAssembly Binaries". In: *International Symposium on Software Testing and Analysis*. ISSTA. New York, NY, USA: Association for Computing Machinery, 2023, pages 385–397. ISBN: 9798400702211. DOI: 10.1145/3597926.3598064.

[33] Bo Jiang, Yifei Chen, Dong Wang, Imran Ashraf, and Wing Kwong Chan. "WANA: Symbolic Execution of Wasm Bytecode for Extensible Smart Contract Vulnerability Detection". In: *Software Quality, Reliability and Security*. QRS. 2021, pages 926–937. DOI: 10.1109/QRS54544.2021.00102.

[34] Mark P. Jones and Luc Duponcheel. *Composing Monads*. Research YALEU/DCS/RR-1004. Department of Computer Science, Yale University, 1993. URL: https://web.cecs.pdx.edu/~mpj/pubs/composing.html (visited on 2024-10-22).

[35] Sarfraz Khurshid, Corina S. Păsăreanu, and Willem Visser. "Generalized Symbolic Execution for Model Checking and Testing". In: *Tools and Algorithms for the Construction and Analysis of Systems*. TACAS. Berlin, Heidelberg: Springer, 2003, pages 553–568. ISBN: 978-3-540-36577-8. DOI: 10.1007/3-540-36577-X_40.

[36] James C. King. "Symbolic execution and program testing". In: *Communications of the ACM* (1976), pages 385–394. DOI: 10.1145/360248.360252.

[37] Daniel Lehmann, Johannes Kinder, and Michael Pradel. "Everything Old is New Again: Binary Security of WebAssembly". In: *USENIX Security Symposium*. SEC. USA: USENIX Association, 2020, pages 217–234. ISBN: 978-1-939133-17-5.

[38] Guodong Li, Esben Andreasen, and Indradeep Ghosh. "SymJS: automatic symbolic testing of JavaScript web applications". In: *Foundations of Software Engineering*. FSE. New York, NY, USA: Association for Computing Machinery, 2014, pages 449–459. ISBN: 978-1-4503-3056-5. DOI: 10.1145/2635868.2635913.

[39] Anil Madhavapeddy and Yaron Minsky. *Real World OCaml: Functional Programming for the Masses*. 2nd edition. Cambridge: Cambridge University Press, 2022. ISBN: 978-1-00-912580-2. DOI: 10.1017/9781009129220.

[40] Filipe Marques, José Fragoso Santos, Nuno Santos, and Pedro Adão. "Concolic Execution for WebAssembly". In: *European Conference on Object-Oriented Programming*. ECOOP. Dagstuhl, Germany: Schloss Dagstuhl – Leibniz-Zentrum für Informatik, 2022, 11:1–11:29. ISBN: 978-3-95977-225-9. DOI: 10.4230/LIPIcs.ECOOP.2022.11.

[41] Brian McFadden, Tyler Lukasiewicz, Jeff Dileo, and Justin Engler. *Security Chasms of WASM*. Technical report. NCC Group, 2018.

[42] Adrian D. Mensing, Hendrik van Antwerpen, Casper Bach Poulsen, and Eelco Visser. "From definitional interpreter to symbolic executor". In: *International Workshop on Meta-Programming Techniques and Reflection*. META. New York, NY, USA: Association for Computing Machinery, 2019, pages 11–20. ISBN: 978-1-4503-6985-5. DOI: 10.1145/3358502.3361269.

[43] Mark Mossberg, Felipe Manzano, Eric Hennenfent, Alex Groce, Gustavo Grieco, Josselin Feist, Trent Brunson, and Artem Dinaburg. "Manticore: A User-Friendly Symbolic Execution Framework for Binaries and Smart Contracts". In: *Automated Software Engineering*. ASE. 2019, pages 1186–1189. DOI: 10.1109/ASE.2019.00133.

[44] Aina Niemetz and Mathias Preiner. "Bitwuzla". In: *Computer Aided Verification*. CAV. Cham: Springer Nature Switzerland, 2023, pages 3–17. ISBN: 978-3-031-37703-7. DOI: 10.1007/978-3-031-37703-7_1.

[45] Louis Noizet and Alan Schmitt. "Semantics in Skel and Necro". In: *Italian Conference on Theoretical Computer Science*. ICTCS. 2022, page 1.

[46] Corina S. Păsăreanu and Neha Rungta. "Symbolic PathFinder: symbolic execution of Java bytecode". In: *Automated Software Engineering*. ASE. New York, NY, USA: Association for Computing Machinery, 2010, pages 179–180. ISBN: 978-1-4503-0116-9. DOI: 10.1145/1858996.1859035.

[47] Andreas Rossberg. *WebAssembly Core Specification*. Technical report. W3C, 2019. URL: https://www.w3.org/TR/wasm-core-1/ (visited on 2024-10-24).

[48] José Fragoso Santos, Petar Maksimović, Théotime Grohens, Julian Dolby, and Philippa Gardner. "Symbolic Execution for JavaScript". In: *International Symposium on Principles and Practice of Declarative Programming*. PPDP. New York, NY, USA: Association for Computing Machinery, 2018, pages 1–14. ISBN: 978-1-4503-6441-6. DOI: 10.1145/3236950.3236956.

[49] Prateek Saxena, Devdatta Akhawe, Steve Hanna, Feng Mao, Stephen McCamant, and Dawn Song. "A Symbolic Execution Framework for JavaScript". In: *IEEE Symposium on Security and Privacy*. SP. 2010, pages 513–528. DOI: 10.1109/SP.2010.38.

[50] Koushik Sen, Darko Marinov, and Gul Agha. "CUTE: a concolic unit testing engine for C". In: *Proceedings of the 10th European Software Engineering Conference Held Jointly with 13th ACM SIGSOFT International Symposium on Foundations of Software Engineering*. ESEC/FSE. New York, NY, USA: Association for Computing Machinery, 2005, pages 263–272. ISBN: 1-59593-014-0. DOI: 10.1145/1081706.1081750.

[51] Koushik Sen, George Necula, Liang Gong, and Wontae Choi. "MultiSE: multi-path symbolic execution using value summaries". In: *Foundations of Software Engineering*. ESEC/FSE. New York, NY, USA: Association for Computing Machinery, 2015, pages 842–853. ISBN: 978-1-4503-3675-8. DOI: 10.1145/2786805.2786830.

[52] Junaid Haroon Siddiqui and Sarfraz Khurshid. "ParSym: Parallel symbolic execution". In: *International Conference on Software Technology and Engineering*. ICSTE. 2010. ISBN: 978-1-4244-8666-3. DOI: 10.1109/ICSTE.2010.5608866.

[53] Julien Signoles, Nikolai Kosmatov, and Kostyantyn Vorobyov. "E-ACSL, a runtime verification tool for safety and security of C programs (tool paper)". In: *International Workshop on Competitions, Usability, Benchmarks, Evaluation, and Standardisation for Runtime Verification Tools*. RV-CuBES. 2017. DOI: 10.29007/FPDH.

[54] The Coq Development Team. *The Coq Reference Manual – Release 8.19.0*. 2024. URL: https://coq.inria.fr/doc/V8.19.0/refman (visited on 2024-10-22).

[55]   Emina Torlak and Rastislav Bodik. "A lightweight symbolic virtual machine for solver-aided host languages". In: *Programming Language Design and Implementation*. PLDI. New York, NY, USA: Association for Computing Machinery, 2014, pages 530–541. ISBN: 978-1-4503-2784-8. DOI: 10.1145/2594291.2594340.

[56]   Guannan Wei, Shangyin Tan, Oliver Bračevac, and Tiark Rompf. "LLSC: a parallel symbolic execution compiler for LLVM IR". In: *Proceedings of the 29th ACM Joint Meeting on European Software Engineering Conference and Symposium on the Foundations of Software Engineering*. ESEC/FSE. New York, NY, USA: Association for Computing Machinery, 2021, pages 1495–1499. ISBN: 978-1-4503-8562-6. DOI: 10.1145/3468264.3473108.

**About the authors**

**Léo Andrès** is a PhD student at OCamlPro and LMF. He is working on compiling OCaml to WasmGC but decided one day to write a Wasm interpreter as a side project.
https://orcid.org/0000-0003-2940-6605

**Filipe Marques** is a PhD Student in Computer Science and Engineering at Instituto Superior Técnico (IST), under the CMU Portugal Affiliated PhD Program. He is also a junior researcher at INESC-ID. His research interests are focused on programming languages, software verification and validation, formal methods, and cybersecurity.
https://orcid.org/0000-0002-2555-5382

**Arthur Carcano** is a research and development engineer at OCamlPro. He likes programming in Rust but sorely misses writing multi-layered monad transformers when doing so.
https://orcid.org/0000-0002-9946-1645

**Pierre Chambart** is CTO and senior research and development engineer at OCamlPro.
https://orcid.org/0009-0008-9163-9091

**José Fragoso Santos** is an Assistant Professor in the Department of Computer Science and Engineering at Instituto Superior Técnico, and a member of INESC-ID.
https://orcid.org/0000-0001-5077-300X

**Jean-Christophe Filliâtre** is a senior researcher at CNRS. He works at LMF and is doing research in deductive program verification.
https://orcid.org/0000-0003-2359-975X